# CCMusic: An Open and Diverse Database for Chinese Music Information Retrieval Research

MONAN ZHOU

SHENYANG XU

ZHAORUI LIU

ZHAOWEN WANG

FENG YU

WEI LI

BAOQIANG HAN

*Author affiliations can be found in the back matter of this article

## ABSTRACT

Data are crucial in various computer-related fields, including music information retrieval (MIR), an interdisciplinary area bridging computer science and music. This paper introduces CCMusic, an open and diverse database comprising multiple datasets specifically designed for tasks related to Chinese music, highlighting our focus on this culturally rich domain. The database integrates both published and unpublished datasets, with steps taken such as data cleaning, label refinement, and data structure unification to ensure data consistency and create ready-to-use versions. We conduct benchmark evaluations for all datasets using a unified evaluation framework developed specifically for this purpose. This publicly available framework supports both classification and detection tasks, ensuring standardized and reproducible results across all datasets. The database is hosted on *HuggingFace* and *ModelScope*, two open and multifunctional data and model hosting platforms, ensuring ease of accessibility and usability.

CORRESPONDING AUTHOR:
**Baoqiang Han**
Department of Musicology, China Conservatory of Music, Beijing, China
hbq@ccmusic.edu.cn

# 1 INTRODUCTION

Music information retrieval (MIR) is a multidisciplinary research field focused on the computational analysis, indexing, retrieval, recommendation, separation, transformation, and generation of music (Müller, 2007). As an interdisciplinary area, modern MIR research heavily relies on machine learning techniques from the field of computer science, particularly, deep learning in recent years. The growing dependence on deep learning has increased the demand for constructing more datasets in the MIR domain.

In the field of MIR, the creation of new datasets has long been a focal point of research. Currently, the research community boasts an extensive array of datasets catering to various tasks, for example, genre classification (Bogdanov et al., 2019), downbeat detection (Hockman et al., 2012), various instruments (e.g., saxophone (Foster and Dixon, 2021) and piano (Hung et al., 2021; Kong et al., 2020)), and different musical genres (e.g., jazz music (Foster and Dixon, 2021) and Western classical music (Thickstun et al., 2017)). However, most of the existing datasets are disseminated within the research community and stored in different data formats. Researchers often need to search extensively to assess the content, reliability, and accessibility of datasets. This process of discovery and examination can be time-consuming and laborious. On this basis, the *mirdata* library (Bittner et al., 2019), which aims at the reproducible usage of MIR datasets, offers an attempt at dataset integration. It provides standardized data loaders for downloading and validating datasets. However, some datasets within it are not freely available. Furthermore, while the library includes a brief guide delineating the annotation types across various datasets to assist users in making informed selections, additional information on dataset contents is not elaborated upon within the document. Additionally, the scarcity of datasets related to Chinese music in this library is noteworthy, reflecting the under-representation of Chinese music within the MIR community.

On the basis of the observations above, we aim to propose a database that encompasses a diverse array of datasets tailored for Chinese-music-related MIR tasks. More importantly, we would like to ensure that all datasets can be searched, viewed, downloaded, and used in one place. Additionally, this database must be openly accessible, eliminating unnecessary barriers for researchers. Inspired by the call for closer collaboration between MIR and music archives (de Valk et al., 2017), this database requires contributions from experts in music and computer-related fields. It is important to clarify that a database is a collection of datasets, and this distinction highlights the scope and organization of the information we aim to provide.

To achieve our goals, we adopt a systematic approach by leveraging our academic networks to consolidate both published and unpublished Chinese-music-related datasets into a unified repository. For unpublished datasets, the importance of integrating and publishing is evident. For published datasets, despite being described in academic papers, many faced challenges in usability, including inconsistent data, partial or missing evaluations, or accessibility barriers. For example, the Chinese Traditional Instrument Sound (CTIS) dataset introduced in Section 3.1 has currently undergone evaluation for only 78 instruments, while the rest are yet to be assessed. Moreover, labels of two datasets are stored in separate metadata files, which increases the complexity of data handling, creating usage obstacles. More importantly, all of the datasets require researchers to submit manual requests and wait for approval before accessing the data, which is an unnecessary impediment to their use. To address these challenges, we consolidate the datasets into a single repository. The integration process such as data structure unification is conducted to standardize aspects such as label–audio associations. Then, we provide statistical information and conduct unified evaluations. To ensure comprehensive evaluation, we employ multiple models and input spectrograms within a unified framework. The database is openly available on *HuggingFace*, allowing researchers to search, view, download, and use them seamlessly in one place, eliminating lengthy application processes. To accommodate researchers affected by regional policy restrictions, we have also made the database available on *ModelScope*, the Chinese equivalent of *HuggingFace*. The contributions of this work can be summarized as follows:

- We collect a series of datasets related to Chinese music, aiming to alleviate the scarcity of Chinese music data in the MIR community.
- For all datasets, we perform a unified integration process, merging them into one unified database. This process involves data cleaning to discard unneeded data, label refinement to remove useless labels or add useful ones, and unifying data structures for better usability. The integrated datasets have been made openly accessible on **HuggingFace**[1] and **ModelScope**[2], ensuring they are fully open and ready to use.
- We conduct a unified evaluation for all datasets using our evaluation framework (which is now open source on **GitHub**[3]). Pre-processed versions of all datasets were produced during evaluation and are now publicly available to save researchers time and improve usability.

The rest of the paper is organized as follows. In Section 2 we provide a review of related work. Section 3 introduces the contents of the database. In Section 4, we present our evaluation framework, along with all the evaluations performed. Finally, Section 5 offers a summary of the paper.

## 2 RELATED WORK

In the field of MIR, many datasets typically contain singular music information (usually tailored for a certain task) such as downbeat in Hockman et al. (2012), tempo in Hainsworth and Macleod (2004) and Levy (2011), genres in Bogdanov et al. (2019), and chords in Burgoyne et al. (2011). Conversely, some datasets encompass multiple types of music information, enabling researchers to address multiple tasks within a single dataset. We present several of these datasets along with their associated labels: MedleyDB (Bittner et al., 2014) (instrument activation, pitch, and genre), the Harmonix set (Nieto et al., 2019) (beat, downbeat, and structure), RWC (Goto, 2006) (rhythm, melody, instrument, structure, genre, etc.), and Music4All (Pegoraro Santana et al., 2020) (key, tempo, and genre). Despite their abundance, datasets are often scattered across the community, with varying data formats. For instance, instrument labels in the RWC dataset are denoted by the folder names storing audio files. In contrast, labels for the Harmonix set are stored in separate metadata files. The fragmented nature of the datasets, with varying data formats and organization, often requires a piecemeal approach to searching, organizing, and utilization, which is time-consuming and tedious.

By integrating the datasets, the problem described above when facing enormous datasets could be alleviated to some extent. The *mirdata* library (Bittner et al., 2019) offers a solution for dataset integration, which aims to address the variability in experimental outcomes caused by dataset version discrepancies, the heterogeneity of annotation file formats, and the diversity of data acquisition methods. It provides standardized interfaces or data loaders for downloading and validating data. It supports 58 datasets at the current stage. Nonetheless, this collection includes 11 datasets that are available upon request and 11 datasets that are not publicly accessible. These non-open data pose an inconvenience for researchers, as the *mirdata* does not host or distribute data. Furthermore, the specifics of each dataset may not be readily presented.

In contrast, in the MIR community, Western music datasets are abundant, encompassing a wide range of data types from audio data (Foster and Dixon, 2021; Thickstun et al., 2017) to MIDI data (Huang et al., 2019; Kong et al., 2020), with many of them being publicly accessible. This abundance facilitates research on Western music and, conversely, encourages the construction of Western music datasets. However, datasets of music from other cultures, such as Chinese music, are comparatively limited. We present all the publicly available datasets that are designed for research topics related to Chinese music in Table 1, highlighting the scarcity of Chinese music in the MIR community. These datasets consist exclusively of traditional Chinese music. In addition to this, there are also datasets related to modern Chinese pop music. However, their creation is aimed at a more general task rather than at topics specifically related to their musical corpus. Under this premise, if the musical corpus in the dataset is replaced with other music, the purpose of the dataset will not be affected. For example, the POP909 (Wang et al., 2020) dataset contains the vocal melody, lead instrument melody, and piano arrangement of Chinese pop music, stored in MIDI format. It was created for the task of music arrangement generation. Furthermore, MIR-1K (Hsu and Jang, 2010) and MIR-ST500 (Wang and Jang, 2021) are two datasets that include singing segments of Chinese pop songs. They were created for the purpose of research in singing voice separation and singing transcription, respectively.

In this work, our database encompasses a diverse range of Chinese music datasets. All associated tasks are closely intertwined with their respective musical corpora. We conduct integration efforts to transform these datasets into unified data structures, significantly enhancing usability. For instance, labels previously stored in separate metadata files are now integrated with the audio data into a single dictionary. Lastly, all datasets are made publicly available, facilitating easy retrieval and use by researchers.

## 3 DATABASE CONTENT

The CCMusic database currently encompasses six datasets: four focused on Chinese musical instruments, one on Chinese music mode, and one on Chinese and Western singing. Table 2 presents the main tasks and main content of these datasets.

In the following sections, we will introduce these datasets in a structured four-step approach: the original content, integration process, statistical analysis, and preparation for evaluation. The integration process for each dataset may involve data cleaning to remove useless data; label refinement, including cleaning up inconsistent labels and adding useful ones such as Chinese pinyin or characters; and data structure unification to standardize storage and improve usability. During evaluation preparation, we convert all audio clips into three commonly used spectrogram types in MIR—mel spectrogram, constant-Q transform (CQT), and chroma—to ensure comprehensive benchmarks. In subsequent sections, we will omit the detailed explanation of the spectrogram transformation process to keep the text concise.

| Dataset | Content | Task | Object |
|---|---|---|---|
| **Corpus of Jingju** (Caro Repetto and Serra, 2014) (2014) | Audio, editorial metadata, lyrics, and scores of Jingju | Melodic analysis | Jingju |
| **SingingDatabase** (Black et al., 2014) (2014) | Vocal recordings by professional, semi-professional, and amateur singers in predominantly Chinese opera singing style | Singing voice analysis | Chinese opera |
| **Unnamed** (Hu and Yang, 2017) (2017) | Traditional Chinese music pieces in the spectrogram representation with information on performers and instruments | Latent space analysis | Chinese traditional music |
| **GQ39** (Huang et al., 2020) (2020) | Audio recordings of prevalent Guqin solo compositions and corresponding event-by-event annotations | Playing technique detection, mode detection | Guqin |
| **Unnamed** (Nahar et al., 2020) (2020) | Musical features of Chinese, Malay, and Indian song fragments | Music classification | Chinese songs |
| **JinYue Database** (Shen et al., 2020) (2020) | Huqin music metadata, audio features, and annotations of emotion, scene, and imagery | Emotion, scene, and imagery recognition | Huqin |
| **ChMusic** (Gong et al., 2021) (2021) | Traditional Chinese music excerpts performed by 11 traditional Chinese musical instruments | Instrument recognition | Chinese instruments |
| **Traditional Chinese Opera** (Zhang et al., 2021) (2021) | Songs from the 14 most popular types of Chinese opera with annotations of music, song, and speech | Genre recognition | Chinese opera |
| **CBFdataset** (Wang et al., 2022a) (2022) | Monophonic recordings of classic Chinese bamboo flute pieces and isolated playing techniques with annotations | Playing technique detection | Zhudi |
| **CCOM-HuQin** (Zhang et al., 2023) (2023) | Audiovisual recordings of 11,992 individual playing technique clips and 57 annotated musical compositions featuring classical excerpts | Playing technique detection | Huqin |

**Table 1** Publicly available datasets designed for topics related to Chinese music in music technology to date.

| Dataset | Main Task | Main Contents |
|---|---|---|
| *Chinese Traditional Instrument Sound* (Liang et al., 2019) | Instrument recognition | Audio of Chinese instruments |
| *GZ IsoTech* (Li et al., 2022) | Playing technique classification | Audio with playing technique annotation |
| *Guzheng Tech99* (Li et al., 2023) | Playing technique detection | Audio with playing technique annotation |
| *Erhu Playing Technique* (Wang et al., 2019) | Playing technique classification | Audio with playing technique annotation |
| *Chinese National Pentatonic Modes* (Wang et al., 2022b) | Mode classification | Audio with Chinese pentatonic mode annotation |
| *Bel Canto & Chinese Folk Singing* (authors' own creation) | Singing style classification | Audio with singing style annotation |

**Table 2** All the datasets included in the CCMusic database (references following the name of each published dataset).

## 3.1 CHINESE TRADITIONAL INSTRUMENT SOUND (CTIS) DATASET
### 3.1.1 Original content
The original dataset was created by Liang et al. (2019), with no evaluation provided. The original Chinese Traditional Instrument Sound (CTIS) dataset contains recordings from 287 varieties of Chinese traditional instruments, reformed Chinese musical instruments, and instruments from ethnic minority groups. Notably, some of these instruments are rarely encountered by the majority of the Chinese populace. The dataset was then utilized by Li and Zhang (2022) for Chinese instrument recognition, where only 78 instruments—approximately one-third of the total instrument classes—were used.

### 3.1.2 Integration

We began by performing data cleaning to remove recordings without specific instrument labels. Additionally, recordings that are not instrumental sounds, such as interview recordings, were removed to enhance usability. Finally, instrument categories lacking specific labels were excluded. The filtered dataset contains recordings of 209 types of Chinese traditional musical instruments. Of the original 287 instrument types, 78 were removed owing to missing instrument labels. Among the remaining instruments, seven have two variants each, and one instrument, Yangqin, has four variants. We treat variants as separate classes, resulting in a total of 219 labels.

In the original dataset, the Chinese character label for each instrument was represented by the folder name housing its audio files. During integration, we added a Chinese pinyin label to make the dataset more accessible to researchers who are not familiar with Chinese. Then, we reorganized the data into a dictionary with five columns, which include audio with a sampling rate of 44,100 Hz, a pre-processed mel spectrogram, a numerical label, the instrument name in Chinese, and the instrument name in Chinese pinyin. The provision of mel spectrograms primarily serves to enhance the visualization of the audio on the *ModelScope* platform. For the remaining datasets, these mel spectrograms will also be included in the integrated data structure. The total data number is 4,956, with a duration of 32.63 hours. The average duration of the recordings is 23.70 seconds.

### 3.1.3 Statistics

Owing to the large number of categories in this dataset, we are unable to provide the audio duration per category and the proportion of audio clips by category, as we have done for the other datasets. Instead, we provide a chart showing the distribution of the number of audio clips across different durations, as shown in Figure 1. A second chart, displayed in Figure 2, shows the instrument category distribution across various durations. From Figure 1, it can be observed that 3,611 clips (73%) are concentrated in the 0–27.5-second range, with a steep drop in the number of samples in longer durations. In Figure 2, about half of the instruments, totaling 117, have a duration of less than 437 seconds, while 102 instruments have a duration greater than this number. After the total duration exceeds 881 seconds, the number of instruments drops sharply. This indicates that the dataset has a certain degree of class imbalance.

### 3.1.4 Evaluation preparation

We first performed silence removal using *librosa* (McFee et al., 2015), ensuring that only meaningful content remains. Then, audio was segmented into 2-second clips. Clips shorter than 2 seconds were circularly padded. Longer clips were cut into 2-second multiples, and the remaining part was circularly padded. At this stage, the audio clip number is 43,054. Finally, for the dataset split, we randomly split the dataset into training, validation, and testing sets at a ratio of 8:1:1.

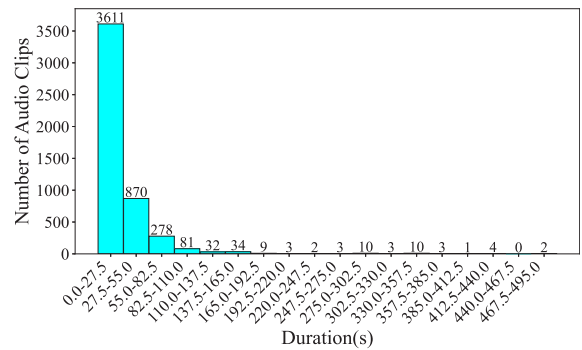

**Figure 1** Number of audio clips across various durations in the CTIS dataset, segmented at 27.5 seconds.

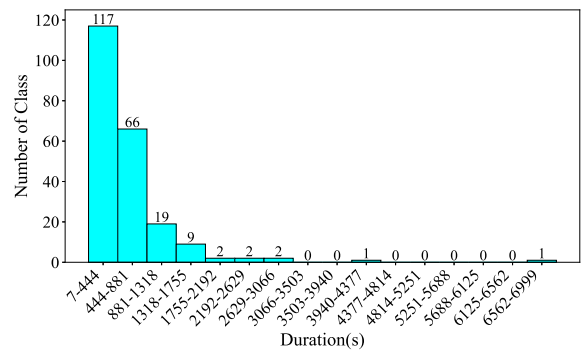

**Figure 2** Number of instrument categories across various durations in the CTIS dataset, segmented at 437 seconds.

## 3.2 GZ ISOTECH DATASET

### 3.2.1 Original content

The dataset was created and used for Guzheng playing technique detection by Li et al. (2022). The original dataset comprises 2,824 variable-length audio clips showcasing various Guzheng playing techniques. Specifically, 2,328 clips were sourced from virtual sound banks, while 496 clips were performed by a professional Guzheng artist. The clips are annotated in eight categories, with Chinese pinyin and Chinese characters written in parentheses: *Vibrato (chanyin* 颤音*), Upward Portamento (shanghuayin* 上滑音*), Downward Portamento (xiahuayin* 下滑音*), Returning Portamento (huihuayin* 回滑音*), Glissando (guazou* 刮奏*, huazhi* 花指…*), Tremolo (yaozhi* 摇指*), Harmonic (fanyin* 泛音*), and Plucks (gou* 勾*, da* 打*, mo* 抹*, tuo* 托…*)*.

### 3.2.2 Integration

In the original dataset, the labels were represented by folder names, which provided Italian and Chinese pinyin labels. During the integration process, we added the corresponding Chinese character labels to ensure comprehensiveness. Lastly, after integration, the data structure has six columns: audio clip sampled at a rate of 44,100 Hz, mel spectrogram, numerical label, Italian label, Chinese character label, and Chinese pinyin label. The number of data after integration remains at 2,824, with a total duration of 63.97 minutes. The average duration is 1.36 seconds.

### 3.2.3 Statistics

Firstly, Figure 3 illustrates the number and proportion of audio clips in each category. The category with the largest proportion is *Upward Portamento*, accounting for 19.0%, with 536 clips. The smallest category is *Tremolo*, accounting for 8.1%, with 228 clips. The difference in proportion between the largest and smallest categories is 10.9%. Next, Figure 4 displays the total audio duration for each category. The category with the longest total duration is *Upward Portamento*, with 14.11 minutes, consistent with the results shown in the pie chart. However, the category with the shortest total duration is *Plucks*, with a total of 5.06 minutes, which differs from the ranking in the pie chart. Overall, this dataset is comparatively balanced within the database. Finally, Figure 5 presents the number of audio clips distributed across different duration intervals. Most of the audio clips are concentrated in the 0–3-second range. Among them, 1,742 audio clips are concentrated in the 0–1.5-second range, while 1,033 clips are concentrated in the 1.5–3-second range. Only 49 clips exceed 3 seconds in length.

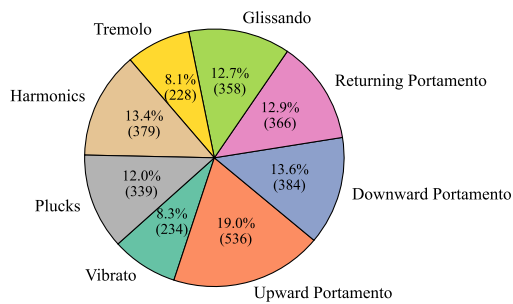

**Figure 3** Clip number and proportion of each category in the GZ IsoTech dataset.

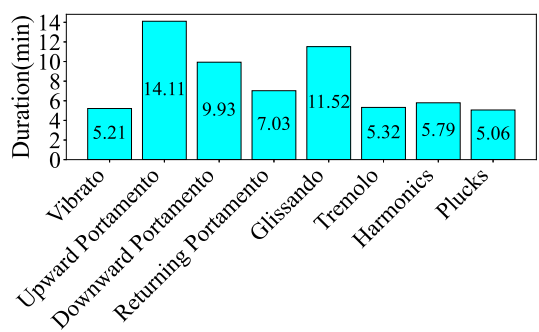

**Figure 4** Audio duration of each category in the GZ IsoTech dataset.

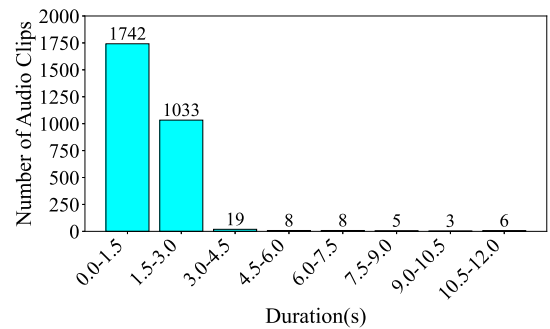

**Figure 5** Number of audio clips across various durations in the GZ IsoTech dataset, segmented at 1.5 seconds.

### 3.2.4 Evaluation preparation

Audio was first segmented into 3-second clips. For clips shorter than 3 seconds, circular padding was applied. Longer clips were cut into 3-second segments, and any remaining parts were circularly padded. At this stage, the audio clip number is 2,899. Finally, the dataset split ratio is 8:1:1.

## 3.3 GUZHENG TECH99 DATASET

### 3.3.1 Original content

This dataset was created and used by Li et al. (2023) for frame-level Guzheng playing technique detection. The original dataset encompasses 99 solo compositions for Guzheng, recorded by professional musicians within a studio environment. Each composition is annotated for every note, indicating the onset, offset, pitch, and playing techniques. This is different from the GZ IsoTech, which is annotated at the clip level. Also, its playing technique categories differ slightly, encompassing a total of seven techniques. They are *Vibrato (chanyin* 颤音*), Plucks (boxian* 拨弦*), Upward Portamento (shanghua* 上滑*), Downward Portamento (xiahua* 下滑*), Glissando (huazhi\guazou\lianmo\liantuo* 花指\刮奏\连抹\连托*), Tremolo (yaozhi* 摇指*),* and *Point Note (dianyin* 点音*)*. This meticulous annotation results in a total of 63,352 annotated labels.

### 3.3.2 Integration

In the original dataset, the labels were stored in a separate CSV file. This posed usability challenges, as researchers had to perform time-consuming operations on CSV parsing and label–audio alignment. After our integration, the data structure has been streamlined and optimized. It now contains three columns: audio sampled at 44,100 Hz, mel spectrogram, and a dictionary. This dictionary contains onset, offset, technique numeric label, and pitch. The number of data entries after integration remains 99, with a cumulative duration amounting to 151.08 minutes. The average audio duration is 91.56 seconds.

### 3.3.3 Statistics

In this part, we present statistics at the label level. The number of audio clips is equivalent to the count of either onset or offset occurrences. The duration of an audio clip is determined by calculating the offset time minus the onset time. At this level, the number of clips is 15,838, and the total duration is 162.69 minutes.

Firstly, Figure 6 illustrates the number and proportion of audio clips in each category. The *Plucks* category accounts for a significantly larger proportion than all other categories, comprising 74.88% of the dataset. The category with the smallest proportion is *Tremolo*, which accounts for only 0.49%, resulting in a difference of 74.39% between the largest and smallest categories. Next, Figure 7 presents the audio duration of each category. The total audio duration of *Plucks* is also significantly longer than all other categories, with a total duration of 117.77 minutes. In contrast, the category with the shortest total duration is *Glissando*, with only 1.13 minutes. This differs from the pie chart, where the smallest category is *Tremolo*. The difference between the longest and shortest durations is 116.64 minutes. From both the pie chart and the duration chart, it is evident that this dataset suffers from a severe data imbalance problem. In the end, Figure 8 gives the audio clip number across various duration intervals. Most of the audio clips are concentrated in the 34–1,004-millisecond range. Among them, 8,660 audio clips fall within the 34–519-millisecond range, 4,935 clips fall within the 519–1,004-millisecond range, and only 2,243 clips exceed 1,004 milliseconds in length.

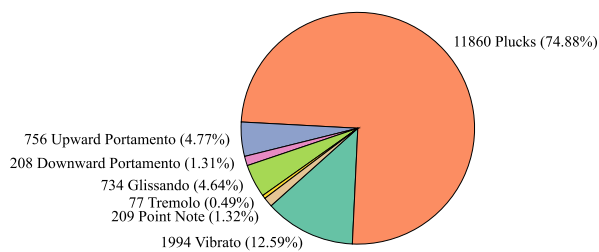

**Figure 6** Clip number and proportion of each category in the Guzheng Tech99 dataset.

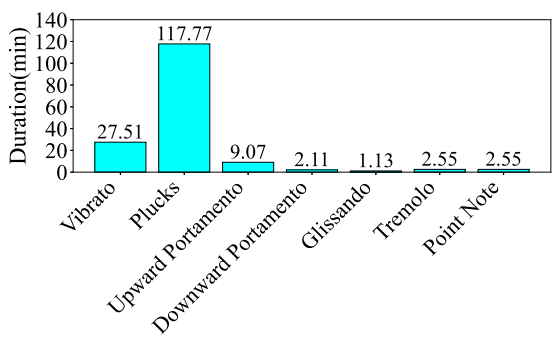

**Figure 7** Audio duration of each category in the Guzheng Tech99 dataset.

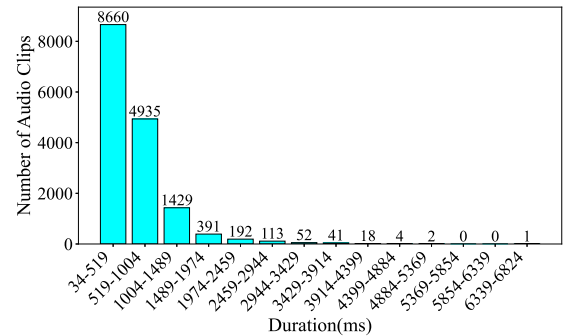

**Figure 8** Number of audio clips across various durations in the Guzheng Tech99 dataset, segmented at 0.485 seconds.

### 3.3.4 Evaluation preparation

For the 99 recordings, silence was first removed on the basis of the annotations, specifically targeting parts without technique annotations. All recordings are uniformly segmented into fixed-length segments of 3 seconds, with clips shorter than 3 seconds being zero-padded. This padding approach, unlike circular padding, is adopted specifically for frame-level detection tasks to prevent the introduction of extraneous information. For the dataset split, since the dataset consists of 99 recordings, we split it at the recording level. The data are partitioned into training, validation, and testing splits in a 79:10:10 ratio, roughly 8:1:1.

## 3.4 ERHU PLAYING TECHNIQUE (ERHUPT) DATASET

### 3.4.1 Original content

This dataset was created and utilized for Erhu playing technique detection by Wang et al. (2019), and has not undergone peer review. The original dataset comprises 1,253 Erhu audio clips, all performed by professional Erhu players. These clips were annotated according to three levels, resulting in annotations for 4, 7, and 11 categories. Part of the audio data is sourced from the CTIS dataset described earlier.

### 3.4.2 Integration

We first performed label cleaning to abandon the labels for the 4 and 7 categories since there are missing data problems. This process left us with only the labels for the 11 categories. Then, we added a Chinese character label and a Chinese pinyin label to enhance comprehensibility. The 11 labels are *Detache (分弓)*, *Diangong (垫弓)*, *Harmonic (泛音)*, *Legato\Slide\Glissando (连弓\滑音\连音)*, *Percussive (击弓)*, *Pizzicato (拨弦)*, *Ricochet (抛弓)*, *Staccato (断弓)*, *Tremolo (震音)*, *Trill (颤音)*, and *Vibrato (揉弦)*.

After integration, the data structure contains six columns: audio (with a sampling rate of 44,100 Hz), mel spectrogram, numeric label, Italian label, Chinese character label, and Chinese pinyin label. The total number of audio clips remains at 1,253, with a total duration of 25.80 minutes. The average duration is 1.24 seconds.

### 3.4.3 Statistics

Figure 9 presents the number of data entries per label. The *Trill* label has the highest data volume, with 249 instances, which accounts for 19.9% of the total dataset. Conversely, the *Harmonic* label has the least amount of data, with only 30 instances, representing 2.4% of the total. Next, regarding the audio duration per category, as illustrated in Figure 10, the audio data associated with the *Trill* label have the longest duration, amounting to 4.88 minutes. In contrast, the *Percussive* label has the shortest audio duration, clocking in at 0.75 minutes. These disparities indicate a class imbalance problem. Finally, as shown in Figure 11, we count the frequency of audio occurrences at 550-millisecond intervals. The quantity of data decreases as the duration lengthens. The most populated duration range is 90–640 milliseconds, with 422 audio clips. The least populated range is 3,390–3,940 milliseconds, which contains only 12 clips.

### 3.4.4 Evaluation preparation

Firstly, we standardized audio clips to a 3-second duration. In this dataset, only few clips exceed 3 seconds, so we circularly padded shorter ones and cut longer ones to 3-second segments, discarding the rest. The audio clip remains at 1,253 at this stage. Regarding the dataset split, given its limited size, we adopt a ratio of 6:2:2 to divide the dataset into training, validation, and testing sets, respectively.

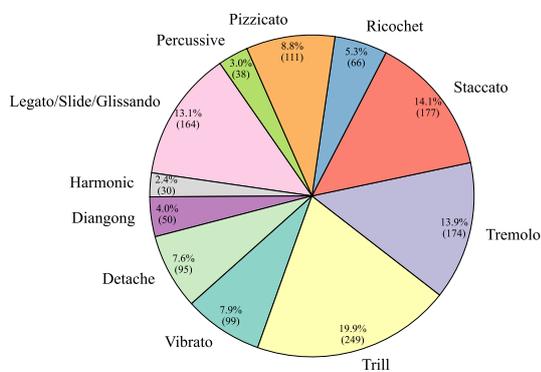

**Figure 9** Clip number and proportion of each category in the ErhuPT dataset.

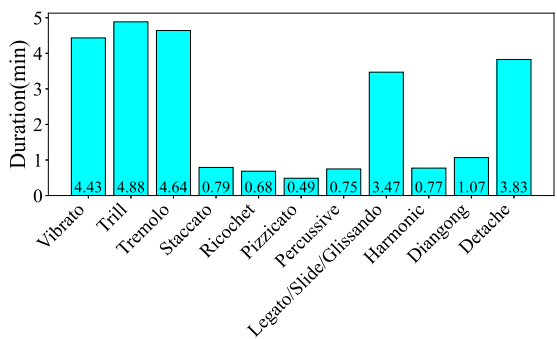

**Figure 10** Audio duration of each category in the ErhuPT dataset.

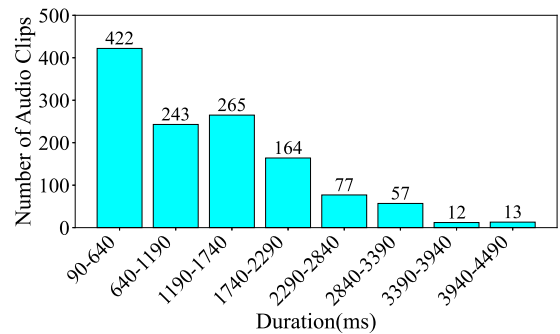

**Figure 11** Number of audio clips across various durations in the ErhuPT dataset, segmented at 550 milliseconds.

## 3.5 CHINESE NATIONAL PENTATONIC MODES (CNPM) DATASET

### 3.5.1 Original content

The dataset was initially created by Ren et al. (2022). It was then expanded and used for automatic Chinese national pentatonic mode recognition by Wang et al. (2022b), to which readers can refer for more details along with a brief introduction to the modern theory of Chinese pentatonic mode. This includes definitions of terms such as "system", "tonic", "pattern", and "type", which will be consolidated into a unified table during our integration process, as described below. The original dataset includes audio recordings and annotations of five modes of Chinese music, encompassing the *Gong (宫), Shang (商), Jue (角), Zhi (徵),* and *Yu (羽)* modes. The total recording number is 287.

### 3.5.2 Integration

Similar to the Guzheng Tech99 dataset in Section 3.3, the labels in this dataset were initially stored in a separate CSV file, which led to certain usability issues. Through our integration, labels were integrated with audio data into a single dictionary. After the integration, the data structure consists of seven columns: The first and the second columns denote the audio recording (sampled at 44,100 Hz) and mel spectrogram. The subsequent columns represent the system, tonic, pattern, and type of the musical piece, respectively. The final column contains an additional Chinese name of the mode. The total number of recordings remains at 287, and the total duration is 858.64 minutes. The average duration is 179.51 seconds.

### 3.5.3 Statistics

In this part, we provide statistics for the "pattern" in the dataset, which includes five categories: *Gong, Shang, Jue, Zhi,* and *Yu*. Our evaluation is also conducted on the basis of "pattern". In a Western context, identifying these patterns is analogous to identifying a musical mode, such as determining whether a piece of music is in the Dorian mode or Lydian mode. These patterns form the core of modern Chinese pentatonic mode theory.

Firstly, Figure 12 presents the number of audio clips by category. *Gong* accounts for the largest proportion among all modes, making up 28.6% of the datasets, with 82 audio clips. The second-largest mode is *Yu*, accounting for 20.6%, with 59 audio clips. The smallest mode is *Shang*, which accounts for 15.7%, with only 45 audio clips. The difference in proportion between the largest and smallest modes is 12.9%. Next, Figure 13 displays the total audio duration by category. The total duration of *Gong* audio is significantly longer than that of other modes, at 290.60 minutes. The second-longest mode is *Yu*, with a total duration of 227.70 minutes, consistent with the proportions shown in the pie chart. However, the shortest mode is not *Shang*, which has the smallest proportion in the pie chart, but rather *Jue*, with a total duration of only 69.13 minutes. The difference in duration between the longest and shortest modes is 221.47 minutes. When we consider the pie chart and the duration statistics together, they expose a data imbalance problem within the dataset. Finally, Figure 14 depicts the number of audio clips across various duration intervals. The interval with the highest number of audio clips is 185–270 seconds, closely followed by 15–100 seconds. Then, once the duration is longer than 355 seconds, the number of audio clips decreases sharply, with only single-digit counts in these longer intervals.

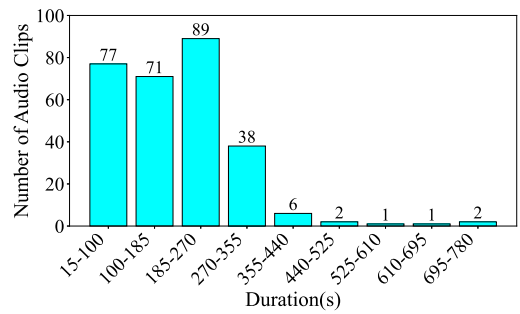

**Figure 14** Number of audio clips across various durations in the CNPM dataset, segmented at 85 seconds.

### 3.5.4 Evaluation preparation
As described above, we classified five "pattern" classes during the evaluation stage. Thus, all clips were segmented into fixed lengths of 20 seconds. The last clip was discarded, as it contains more information about music mode than other clips. After pre-processing, the number of clips is 2,724. In the end, the dataset is split in a ratio of 8:1:1.

## 3.6 BEL CANTO & CHINESE FOLK SINGING DATASET
### 3.6.1 Original content
This dataset was created by the authors. It encompasses two distinct singing styles: bel canto and Chinese folk singing. Bel canto is a vocal technique frequently employed in Western classical music and opera, symbolizing the zenith of vocal artistry within the broader Western musical heritage. We opt to use the term "Chinese folk singing", which lacks an official English translation, to denote a classical singing style that originated in China during the 20th century. It is a fusion of traditional Chinese vocal techniques with European bel canto singing, and it is currently widely utilized in the performance of various forms of Chinese folk music. The purpose of creating this dataset is to fill a gap in existing singing datasets, as none of them include Chinese folk singing. Moreover, by incorporating both bel canto and Chinese folk singing, it provides a valuable resource for researchers to conduct cross-cultural comparative studies in vocal performance.

The original dataset contains 203 a capella singing recordings that are sung in two styles, bel canto and Chinese folk singing. All of them are sung by professional vocalists and were recorded in the recording studio of the China Conservatory of Music using a Schoeps MK4 microphone. Additionally, apart from singing style label, a gender label is also included.

### 3.6.2 Integration
Since this is a self-created dataset, we directly carried out the unified integration of the data structure. After integration, the data structure of the dataset is as follows: audio (with a sampling rate of 22,050 Hz), mel spectrogram, four-class numerical label, gender label, and singing style label. The data number remains 203, with a total duration of 5.08 hours. The average duration is 90 seconds.

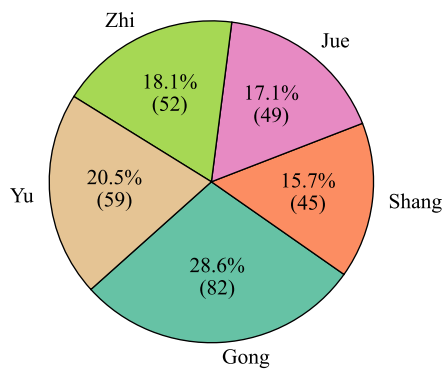

**Figure 12** Clip number and proportion of each category in the CNPM dataset.

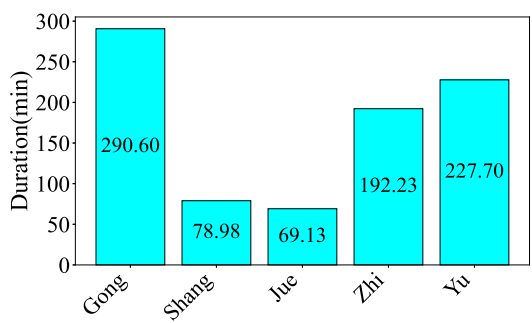

**Figure 13** Audio duration of each category in the CNPM dataset.

### 3.6.3 Statistics

Firstly, Figure 15 presents the clip number for each category. The label with the highest data volume is *Folk Singing Female*, comprising 93 audio clips, which is 45.8% of the dataset. The label with the lowest data volume is *Bel Canto Male*, with 32 audio clips, constituting 15.8% of the dataset. Next, we assessed the length of audio for each label in Figure 16. The *Folk Singing Female* label has the longest audio data, totaling 119.85 minutes. *Bel Canto Male* has the shortest audio duration, amounting to 46.61 minutes. This trend in audio duration aligns with the distribution of data entries shown in the pie chart. Lastly, in Figure 17, the number of audio clips within various duration intervals is displayed. The most common duration range is observed to be 46–79 seconds, with 78 audio segments, while the least common range is 211–244 seconds, featuring only 2 audio segments.

### 3.6.4 Evaluation preparation

Firstly, silence removal was performed. Then, we segmented the audio into fixed-length slices, each 1.6 seconds long. The head and tail slices that may not be representative were discarded. After this process, 9,910

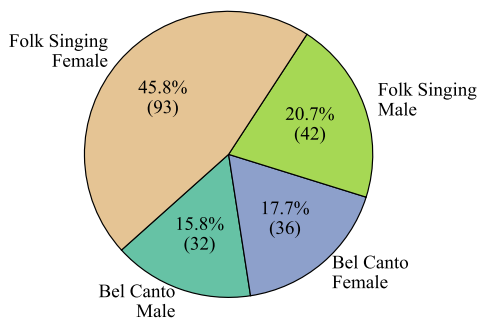

**Figure 15** The clip number and proportion of each category in the Bel Canto & Chinese Folk Singing dataset.

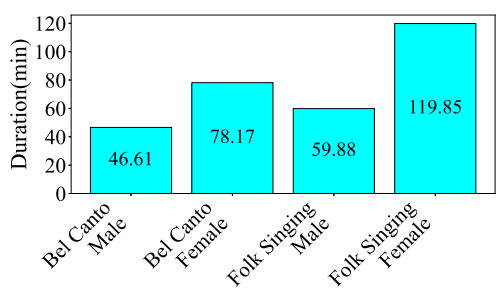

**Figure 16** Audio duration of each category in the Bel Canto & Chinese Folk Singing dataset.

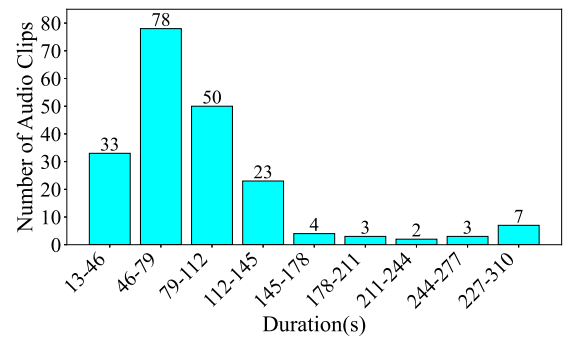

**Figure 17** Number of audio clips across various durations in the Bel Canto & Chinese Folk Singing dataset, segmented at 33 seconds.

slices were obtained. In the end, the dataset is split in a ratio of 8:1:1.

## 4 EVALUATIONS

In this section, we perform benchmark evaluations on all datasets. Within our database, five datasets are designed for classification tasks, while Guzheng Tech99 is designed for a frame-level detection task. To ensure consistency, we developed a reusable evaluation framework that supports both classification and frame-level detection, allowing for standardized evaluations across all datasets. In the following sections, we first introduce the evaluation framework, then describe the experimental setup, and finally present the evaluation results for all datasets.

### 4.1 THE EVALUATION FRAMEWORK

Our framework is inspired by the works of Palanisamy et al. (2020) and Tsalera et al. (2021), both of which approached audio classification as an image classification task in the field of computer vision (CV) and utilized convolutional neural networks (CNN) pre-trained on image data, achieving promising results. Building on this method, our framework utilizes a series of image classification networks from the CV domain as backbones. These networks are pre-trained on ImageNet, and their original classification heads are replaced with custom classifier and detector heads for audio classification and frame-level detection.

Our framework, as illustrated in Figure 18, began with dataset pre-processing, where audio clips are converted into three commonly used spectrogram representations in MIR: mel, CQT, and chroma. The backbone models and their pre-trained weights were obtained from *PyTorch*. Next, the input spectrograms were resized to match the requirements of the selected network before being passed through the backbone. The backbone's final module (e.g., classifier) was replaced with two custom-designed heads. The classifier is tailored for audio classification tasks, and the detector is designed for frame-level detection tasks.

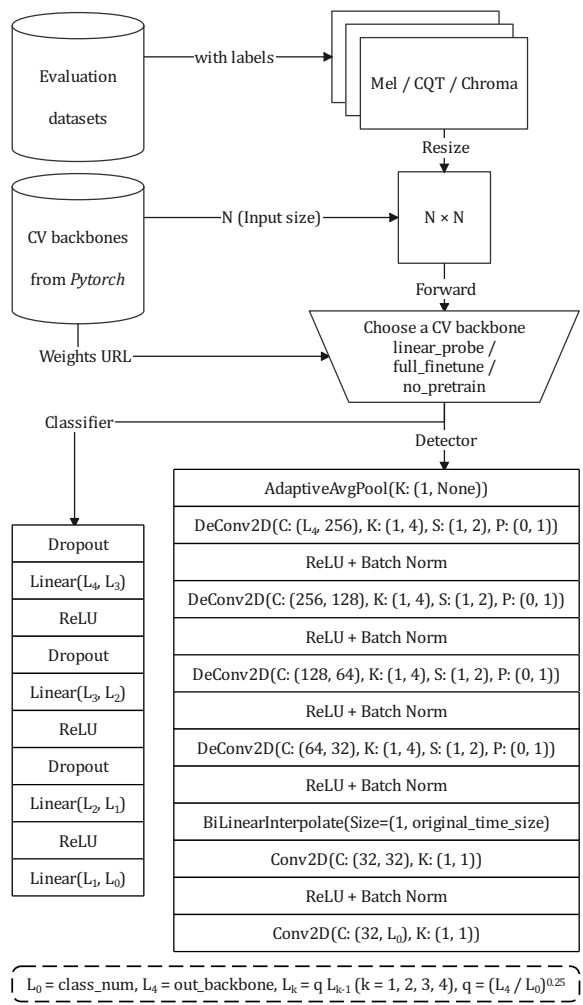

**Figure 18** The evaluation framework that supports classification and detection tasks.

The classifier head has a simple yet effective design. It consists of four linear layers, with a geometric progression in the number of neurons in each layer. The common ratio between the number of neurons in consecutive layers can be deduced on the basis of the output size of the backbone and the number of classification categories. The number of neurons in each linear layer was then arranged by rounding down the computed common ratio. To prevent potential issues such as gradient vanishing and overfitting during training, we incorporated dropout between adjacent linear layers.

The detector head is designed to meet the specific requirements of frame-level detection, ensuring that the output feature vector retains the time dimension. This contrasts with standard classification networks, which typically compress both the frequency and time dimensions. To ensure compatibility with frame-level detection, the detector head processed the backbone's output through a series of steps, upscaling the time dimension and reducing the channel dimension to align with the number of classes. First, the feature output by the backbone underwent average pooling along the frequency dimension, reducing it to one dimension. Next, the time dimension was upsampled using four deconvolutional layers, increasing its resolution and reducing the channel dimension to 32. To standardize the time dimension across varying backbone outputs, a bilinear interpolation resized the time dimension to match the original size of the input spectrogram. Finally, two convolutional layers transformed the channel dimension to the number of classes, producing the output required for frame-level detection. This design ensures that the time resolution is preserved and appropriately processed for detection tasks.

The list of supported neural network backbones and weights is presented in the "Classification" section of the *PyTorch*[4] website, with a total of 18 models. Of these, 3 are transformer models, and 15 are CNN models.

### 4.2 EXPERIMENTAL SETUP

First, we selected several models on the basis of the size of the datasets and the number of parameters in each model. Then, we ran two transformer-based models and five CNN models. In the next experimental results section, we report the outcomes for these models, totaling the results of seven models.

All the selected models were pre-trained on ImageNet V1 (Deng et al., 2009) and then fully fine-tuned using a stochastic gradient descent optimizer with an initial learning rate of 0.001 for 40 epochs. The batch size was set to 4. To address varying degrees of data imbalance across six datasets, we used weighted cross-entropy as the loss function for classification tasks and weighted binary cross-entropy for detection tasks. Then, we adopted the F1-score as the primary metric, as it is well-suited to assessing model performance in the presence of imbalanced datasets.

### 4.3 RESULTS

In this part, we present the evaluation results for the six datasets of our database. Within each result table, models are grouped into transformer and CNN categories and are listed in descending order of parameter size.

#### 4.3.1 CTIS dataset

The classification results of the CTIS dataset are presented in Table 3. The comparison between the transformer and CNN groups highlights that the CNN group achieves superior performance overall. Specifically, the best-performing model is RegNet-Y-32GF, a CNN model, which achieves an F1-score of 0.980 when using CQT as input. This score outperforms the best result from the transformer group by a margin of 0.024. The top-performing transformer model, Swin-T, achieves an F1-score of 0.956 using mel spectrograms as input.

A clear trend emerges when examining the impact of model size within the CNN group. The best-performing model, RegNet-Y-32GF, is also the largest model. Conversely, the smallest CNN model, Inception-V3, performed the worst among all models, with F1-scores below 0.9 across all input types. This suggests that the number of parameters plays a critical role in achieving high performance. In the transformer group, Swin-T, despite being smaller than ViT-L-32, achieved better results across two inputs: mel spectrograms and CQT. This suggests that architecture design and efficiency play a more critical role in the transformer group compared with sheer parameter count.

| Backbone | Mel | CQT | Chroma |
|---|---|---|---|
| ViT-L-32 | 0.936 | 0.921 | **0.845** |
| Swin-T | **0.956** | **0.940** | 0.759 |
| RegNet-Y-32GF | **0.973** | **0.980** | 0.848 |
| VGG19-BN | 0.966 | 0.965 | **0.852** |
| AlexNet | 0.936 | 0.921 | 0.661 |
| ResNet101 | 0.953 | 0.949 | 0.782 |
| Inception-V3 | 0.860 | 0.855 | 0.664 |
| Average | 0.940 | 0.933 | 0.773 |

**Table 3** F1-scores of seven models on the CTIS dataset. The best results of the transformer group and the CNN group are indicated in bold.

| Backbone | Mel | CQT | Chroma |
|---|---|---|---|
| ViT-L-16 | **0.855** | **0.824** | **0.770** |
| MaxVit-T | 0.763 | 0.776 | 0.642 |
| ResNeXt101-64X4D | 0.713 | 0.765 | 0.639 |
| ResNet101 | 0.731 | 0.798 | **0.719** |
| RegNet-Y-8GF | 0.804 | **0.807** | 0.716 |
| ShuffleNet-V2-X2.0 | 0.702 | 0.799 | 0.665 |
| MobileNet-V3-Large | **0.806** | 0.798 | 0.657 |
| Average | 0.768 | 0.795 | 0.687 |

**Table 4** F1-scores of seven models on the GZ IsoTech dataset. The best scores of the transformer group and the CNN group are indicated in bold.

Regarding input representations, mel spectrograms consistently provide robust results, with the best average F1-score, 0.940, narrowly outperforming CQT by 0.007. However, CQT achieved the best overall result, which highlights that it is particularly effective for certain high-capacity models.

In summary, CNN models are better suited for this dataset, with RegNet-Y-32GF achieving the highest F1-score, 0.980, using CQT. Larger models with more parameters tend to perform better, highlighting the importance of model capacity for extracting meaningful features. Among input types, mel spectrograms provide the best average results, while the CQT spectrogram delivers the best individual performance.

### 4.3.2 GZ IsoTech dataset

The classification results of the GZ IsoTech dataset are presented in Table 4. The transformer group shows relatively higher performance than the CNN group, particularly with ViT-L-16. It achieved the highest F1-score across all configurations, reaching 0.855 on the mel spectrogram, highlighting its capacity to effectively model global dependencies within the dataset. In the CNN group, the best performance was achieved by RegNet-Y-8GF on CQT input, with an F1-score of 0.807. This result is 0.048 points lower than ViT-L-16's top performance, underscoring the dominance of transformers in this task. However, it is noteworthy that MobileNet-V3-Large achieved nearly identical performance (0.806) on mel spectrograms, despite being a much smaller model compared with RegNet-Y-8GF. This highlights the efficiency of lightweight architectures in certain configurations. Interestingly, the larger ResNeXt101-64X4D and ResNet101 models underperformed relatively to smaller or mid-sized CNNs, such as MobileNet and RegNet. This suggests that excessively large CNNs may suffer from overfitting in this dataset.

Among the three input types, CQT emerged as the best-performing feature overall, with an average F1-score of 0.795. It outperformed mel spectrograms (average F1 of 0.768) and chroma (average F1 of 0.687). Despite this, mel spectrograms enabled the single highest score (0.855 by ViT-L-16), suggesting that specific models may better exploit certain input types.

The analysis reveals that transformer models, particularly ViT-L-16, dominate in terms of performance across different input types, making them a compelling choice for this dataset. While CNNs such as RegNet-Y-8GF and MobileNet-V3-Large generally lag behind transformers, they offer competitive alternatives with smaller computational costs. Among all input types, CQT offers the highest average performance, though mel spectrograms achieved the highest single score when paired with the right architecture.

### 4.3.3 Guzheng Tech99 dataset

The detection results of the Guzheng Tech99 datasets are shown in Table 5. Unlike the GZ IsoTech dataset, where transformer models dominated, the CNN group achieved superior performance for this dataset. Specifically, VGG19, the largest CNN model in the group, achieved the highest F1-score, 0.862, using mel spectrograms as input. This result highlights its ability to capture and leverage localized features in detection tasks effectively. In the transformer group, Swin-T achieved the best performance, with an F1-score of 0.849 using mel spectrograms. This score is 0.013 points lower than the top-performing CNN model, indicating that, while transformers remain competitive, CNNs maintain a slight edge on this dataset. Notably, ConvNeXt-B from the CNN group also achieved an F1-score of 0.849 on both mel and CQT, matching Swin-T on the mel spectrogram. This further reinforces the competitive strength of CNNs in this dataset. The difference in model group dominance between this dataset and the GZ IsoTech might be attributed to the nature of detection tasks. Unlike the classification task in GZ IsoTech, the Guzheng Tech99 dataset involves a frame-level detection task. CNNs excel at capturing local and fine-grained features, which likely gives them an advantage in this scenario.

| Backbone | Mel | CQT | Chroma |
|---|---|---|---|
| ViT-B-16 | 0.705 | 0.518 | 0.508 |
| Swin-T | **0.849** | **0.783** | **0.766** |
| VGG19 | **0.862** | 0.799 | 0.665 |
| EfficientNet-V2-L | 0.783 | 0.812 | 0.697 |
| ConvNeXt-B | 0.849 | **0.849** | **0.805** |
| ResNet101 | 0.638 | 0.830 | 0.707 |
| SqueezeNet1.1 | 0.831 | 0.814 | 0.780 |
| Average | 0.788 | 0.772 | 0.704 |

**Table 5** F1-scores of seven models on the Guzheng Tech99 dataset. The best scores of the transformer group and the CNN group are indicated in bold.

| Backbone | Mel | CQT | Chroma |
|---|---|---|---|
| Swin-S | 0.978 | 0.940 | 0.903 |
| Swin-T | **0.994** | **0.958** | **0.957** |
| AlexNet | 0.960 | 0.970 | 0.933 |
| ConvNeXt-T | **0.994** | **0.993** | **0.954** |
| ShuffleNet-V2-X2.0 | 0.990 | 0.923 | 0.887 |
| GoogleNet | 0.986 | 0.981 | 0.908 |
| SqueezeNet1.1 | 0.932 | 0.939 | 0.875 |
| Average | 0.976 | 0.958 | 0.917 |

**Table 6** F1-scores of seven models on the ErhuPT dataset. The best scores of the transformer group and the CNN group are indicated in bold.

When analyzing performance across the input types, the mel spectrogram emerged as the most effective feature representation, with the highest average F1-score, 0.788. This was followed closely by CQT, with an average F1-score of 0.772, and chroma, which trailed at 0.704. These trends suggest that the mel spectrogram provides richer and more discriminative features for this dataset. Also, the highest result, which is 0.862, is also on mel spectrogram, further emphasizing its effectiveness.

In summary, VGG19 with mel spectrogram input achieved the best overall result, demonstrating the suitability of CNNs for this dataset. While transformers performed well, particularly Swin-T with mel input, CNNs exhibited a slight performance advantage for this task. Then, among all input types, mel spectrograms demonstrated the strongest average performance, making them the most effective input type for this dataset.

### 4.3.4 ErhuPT dataset

The classification results of the ErhuPT dataset are shown in Table 6. Initially, ConvNeXt-T and Swin-T both achieved the highest scores, 0.994, when the input was mel spectrograms. These two models also deliver strong and comparable results for the other input types, CQT and chroma, reinforcing their robustness across different input representations. Notably, while ConvNeXt-T and Swin-T led in terms of performance, other models such as GoogleNet (0.986 on mel spectrogram) and AlexNet (0.960 on mel spectrogram) also achieved strong results, showing that even classic architectures can remain competitive in this dataset. Lightweight models such as ShuffleNet-V2-X2.0 (0.990 on mel) and SqueezeNet1.1 (0.932 on mel) performed well relative to their size, offering efficient alternatives with slightly reduced performance.

The analysis of input types reveals that mel spectrograms provided the most comprehensive representation for distinguishing the nuanced playing techniques of this Erhu dataset. With an average F1-score of 0.976, mel spectrograms outperformed CQT, which had an average score of 0.958, and chroma features, which achieved the lowest average score of 0.917.

In conclusion, mel spectrograms have been shown to be the most effective input representation for this dataset, not only enabling the highest model performance (ConvNeXt-T and Swin-T, both achieving 0.994) but also achieving the highest overall average performance (0.976). This highlights the strength of mel spectrograms in providing rich and detailed time–frequency information essential for this dataset. Additionally, it is noteworthy that the evaluation results for this dataset are comparatively high, with the average score across all three input types exceeding 0.9.

### 4.3.5 CNPM dataset

The classification results of the CNPM dataset are presented in Table 7. The transformer group demonstrated the most promising performance, particularly with the ViT-L-16 model. This model achieved the highest

| Backbone | Mel | CQT | Chroma |
|---|---|---|---|
| ViT-L-32 | 0.680 | 0.769 | 0.399 |
| ViT-L-16 | **0.823** | **0.859** | **0.549** |
| VGG11-BN | **0.807** | **0.843** | **0.609** |
| RegNet-Y-16GF | 0.590 | 0.832 | 0.535 |
| Wide-ResNet50-2 | 0.694 | 0.757 | 0.531 |
| AlexNet | 0.742 | 0.744 | 0.542 |
| ShuffleNet-V2-X2.0 | 0.473 | 0.720 | 0.266 |
| Average | 0.687 | 0.789 | 0.490 |

**Table 7** F1-scores of seven models on CNPM dataset. The best scores of the transformer group and the CNN group are indicated in bold.

| Backbone | Mel | CQT | Chroma |
|---|---|---|---|
| Swin-S | **0.928** | **0.936** | **0.787** |
| Swin-T | 0.906 | 0.863 | 0.731 |
| AlexNet | 0.919 | 0.920 | 0.746 |
| ConvNeXt-T | 0.895 | 0.925 | 0.714 |
| GoogleNet | **0.948** | 0.921 | 0.739 |
| MNASNet1.3 | 0.931 | **0.931** | **0.765** |
| SqueezeNet1.1 | 0.923 | 0.914 | 0.685 |
| Average | 0.921 | 0.916 | 0.738 |

**Table 8** F1-scores of seven models on the Bel Canto & Chinese Folk Singing dataset. The best scores of the transformer group and the CNN group are indicated in bold.

F1-score, 0.859, when using the CQT input and also excelled with the mel spectrogram input, attaining an F1-score of 0.823. These results underline the robustness of the ViT-L-16 architecture across different input representations. In contrast, the ViT-L-32 model, which has a larger parameter count compared with ViT-L-16, exhibited a noticeable decline in performance. The performance drop may indicate that factors such as overfitting or optimization difficulties might play a role when using larger transformer architectures. Within the CNN group, the VGG11-BN model emerged as the top performer, with an F1-score of 0.843 using the CQT input. However, its performance was slightly behind the ViT-L-16 by a margin of 0.016, underscoring the competitive edge of transformer models on this dataset. In contrast, ShuffleNet-V2-X2.0, the smallest CNN model, recorded the lowest performance across all spectrogram types, particularly with the chroma input, where it achieved an F1-score of just 0.266. Its underperformance suggests that model size and architecture play a crucial role in achieving effective results for complex mode detection tasks.

When analyzing the average performance across the three input types, CQT emerged as the most effective representation, achieving the highest average F1-score, 0.789. This surpasses the mel spectrogram's average of 0.687 by 0.102, further solidifying CQT as the most robust input type for this dataset.

In conclusion, while transformers such as ViT-L-16 show state-of-the-art results, certain CNN models such as VGG11-BN remain competitive in specific contexts. Then, CQT representation proves to be the most effective input overall, achieving the highest average performance, while chroma lags significantly.

### 4.3.6 Bel Canto & Chinese Folk Singing dataset

The classification results of this dataset are presented in Table 8. GoogleNet achieved the highest F1-score, 0.948, with mel spectrogram input, while the top transformer model, Swin-S, reached 0.936 with CQT input. Notably, GoogleNet not only achieved the highest overall score but also delivered a consistently strong performance across all input types, highlighting its versatility and robustness. Moreover, within the CNN group, both GoogleNet and MNASNet1.3 demonstrated stable and high performance, particularly with mel spectrogram and CQT inputs, further solidifying their positions as reliable architectures for this dataset.

Regarding input types, mel spectrograms emerged as the best-performing representation overall, achieving an average F1-score of 0.921 across all models. CQT inputs followed closely, with an average score of 0.916, proving their effectiveness as a viable alternative. However, chroma features significantly underperformed, with an average F1-score of 0.738, likely due to their limited ability to represent timbral and harmonic information critical for distinguishing singing styles in this dataset.

In conclusion, GoogleNet within the CNN group outperformed Swin-S in the transformer group, reaffirming its adaptability and strength across different input types. Among all input types, the mel spectrogram was the most effective spectral input for this benchmark evaluation, with CQT offering competitive performance, particularly for transformer models.

## 5 SUMMARY

In this work, we presented CCMusic, an open and diverse database designed specifically for MIR research related to Chinese music, mitigating the lack of sufficient Chinese music datasets in the MIR community. The database integrates both published and unpublished datasets. Through a unified process involving data cleaning, label refinement, and standardization of data structures, we have prepared ready-to-use versions of these datasets. The consolidated datasets are openly accessible on platforms such as *HuggingFace* and *ModelScope*, enabling researchers to search, view, download, and use the datasets conveniently without the need for cumbersome application procedures. We performed a unified evaluation of all datasets using a self-developed evaluation framework, which supports both classification and frame-level detection tasks. The framework, which is publicly available on *GitHub*, ensures reproducibility and saves researchers time by providing pre-processed versions of the datasets. Overall, CCMusic provides a valuable resource on Chinese music for the MIR community by consolidating diverse datasets, enhancing usability, and offering unified benchmarks.

## ETHICS AND CONSENT

All data in this database were obtained from human participants, and we have received their oral informed consent. Additionally, all data have been anonymized to protect the participants' privacy.

## ACKNOWLEDGEMENTS

We thank Yuan Wang for contributing computational resources. We also thank Dichucheng Li and Yulun Wu for their discussion regarding the detail alignment in the Guzheng datasets.

## FUNDING INFORMATION

This work was supported by the National Social Science Fund of China (Grant No. 22VJXG012 and No. 21ZD19), and the Special Program of the National Natural Science Foundation of China (Grant No. T2341003).

## COMPETING INTERESTS

The authors have no competing interests to declare.

## AUTHOR CONTRIBUTIONS

Monan Zhou and Shenyang Xu made equal contributions to this work.

## NOTES

1. https://huggingface.co/collections/ccmusic-database/ccmusic-benchmarks-66af4edb42c34e7a21e2d163
2. https://www.modelscope.cn/collections/CCMUSIC-lunwenjixian-1632d8d8fabb41
3. https://github.com/monetjoe/ccmusic_eval
4. https://pytorch.org/vision/main/models

## AUTHOR AFFILIATIONS

**Monan Zhou**  https://orcid.org/0009-0001-7317-0727
Department of Music AI and Information Technology, Central Conservatory of Music, Beijing, China

**Shenyang Xu**  https://orcid.org/0000-0002-2347-6939
Department of Music AI and Information Technology, Central Conservatory of Music, Beijing, China

**Zhaorui Liu**  https://orcid.org/0009-0005-7859-5918
Department of Musicology, China Conservatory of Music, Beijing, China

**Zhaowen Wang**  https://orcid.org/0000-0003-4761-5363
Department of Music AI and Information Technology, Central Conservatory of Music, Beijing, China

**Feng Yu**
Department of Music AI and Information Technology, Central Conservatory of Music, Beijing, China

**Wei Li**  https://orcid.org/0000-0002-4486-8341
School of Computer Science and Technology, Fudan University, Shanghai, China; Shanghai Key Laboratory of Intelligent Information Processing, Fudan University, Shanghai, China

**Baoqiang Han**
Department of Musicology, China Conservatory of Music, Beijing, China